\documentclass[traditabstract]{aa}
\usepackage{graphicx}
 \usepackage{lscape}
 \usepackage{colortbl}
 \usepackage{natbib}
 \usepackage{verbatim} 
  \usepackage{comment} 
\usepackage{ulem}
\usepackage{url}
 \bibpunct{(}{)}{;}{a}{}{,}
\usepackage{txfonts}
\begin{document}
\input psfig.tex
\def\simgt{\stackrel{>}{{}_\sim}}
\def\simlt{\stackrel{<}{{}_\sim}}

\titlerunning{$z\sim1.5$ mass function of galaxies in clusters}

\title
{Observational evidence that massive cluster galaxies were
forming stars at $z\sim2.5$ and did not grow in mass at later times}
\author{S. Andreon} 
\institute{INAF-Osservatorio Astronomico di Brera, via Brera 28, 20121, Milano, Italy\\
\email{stefano.andreon@brera.inaf.it} 
}
\date{Received --, 2012; accepted --, 2012}

\abstract{Using Spitzer 3.6 micron data we derived the luminosity function
and the mass function of galaxies in five 
$z>1.4$ clusters selected to have a firm intracluster medium detection. 
The five clusters differ in richness (ISCS\,J1438.1+3414 and 
XMMXCS\,J2215.9-1738 are twice as rich as ISCS\,J1432.4+3250, 
IDCS\,J1426.5+3508, and JKCS\,041) and morphological appareance.
At the median redshift
$z=1.5$, from
the 150 member galaxies of the five clusters,
we derived a characteristic magnitude of $16.92\pm0.13$ in the
[3.6] band and a 
characteristic mass of $lgM^*=11.30\pm 0.05$ M$_\odot$. 
We find
that the characteristic luminosity and mass does not evolve 
between $z=1$ and $1.4<z<1.8$, directly ruling out ongoing mass
assembly between these epochs because massive galaxies are already
present up to $z=1.8$. Lower--redshift build--up epochs have
already been ruled out by previous works, leaving only $z>1.8$ as a possible
epoch for the mass build up. However,
the observed values of $m^*$ at very
high redshift are too bright for galaxies without any
star formation immediately preceding the observed redshift
and therefore
imply a star formation episode not earlier than $z_f=2.5$. 
For the first time, mass/luminosity
functions are able to robustly distinguish tiny differences between
formation redshifts and to set
{\it upper} limits to the epoch of the last star-formation episode.
}

   \maketitle

\section{Introduction}

The galaxy mass--assembly history 
can be reconstructed via the infrared luminosity function (LF), which is
sensitive to the growth of the stellar mass of galaxies as a function of
time. 
The available data, almost entirely at $z<1.2$, are usually interpreted as
evidence that that most of the bright galaxies
must have also been largely assembled by this
redshift (De Propris et al. 1999;  Andreon
2006; De Propris et al. 2007; Muzzin et al. 2008, Andreon et al. 2009).
Andreon (2006) was the first work to sample the
high redshift range well (six clusters above $z=0.99$), and it excluded
several mass growth models, in particular a twofold increase in mass 
over a 8 Gry period.
While consistent
with a scenario where galaxies in clusters are fully assembled at high
redshift, later works (e.g. De Propris et al.
2007; Muzzin et al. 2008; Strazzullo et al. 2010), are unable to give a more stringent constraint, 
mainly because 
a large redshift baseline and precise $m^*$ are needed
to distinguish scenarios, whereas very few
clusters at high redshift were known, and almost none was present 
in the studied
samples. 

At variance with these works, Mancone et al. (2010) studied
candidate clusters, i.e. cluster detections, and claim to have finally
reached the epoch of galaxy mass assembly. This epoch is at 
$z\approx1.3$, because $m^*$ is much fainter at $z\ga1.3$ than it
should be for galaxies if there is no mass growth. We are, according to 
these authors, seeing the rapid mass assembly of cluster galaxies at $z\approx1.3$.
However, this result is in tension with 
the K-band LF of the $z=1.39$ cluster 
1WGA\,J2235.3-2557 (Strazzullo et al. 2010), with
the existence of
a developed and tight red sequence at even higher redshifts, as high as 
$z=1.8$ (Andreon \& Huertas-Company 2011, Andreon 2011), and with a very
early quenching of the cluster populations (Raichoor \& Andreon 2012b). 
All this evidence implies
a peacefully evolution for cluster galaxies up to $z=1.8$.

Since a number of ``bona fide" clusters
at high redshift have been discovered  in the past very few years 
and  for many of them deep
Spitzer observations are 
available, we decided to measure the galaxy mass function of 
$z>1.4$ clusters. 

Throughout this paper, we assume $\Omega_M=0.3$, $\Omega_\Lambda=0.7$, 
and $H_0=70$ km s$^{-1}$ Mpc$^{-1}$. Magnitudes are quoted in their
native system (Vega for I and [3.6] bands, AB for the z' band).  
Unless otherwise stated, results of the statistical computations 
are quoted in the form $x\pm y$, where $x$
is the posterior mean and $y$ the posterior standard deviation.

\section{The sample and the data}

\subsection{Sample}

In this work we  studied the luminosity and mass
function of galaxies in $z>1.4$ clusters 
with a firm detection of the intracluster medium (ICM).
We adopted this choice not to run the risk of including
in the cluster sample other high redshift structures
that have been named ``cluster", but whose
nature is uncertain, or just different (e.g. a proto-cluster).
Since we are interested in a clean sample of secure clusters, we 
applied a severe screening to the list of objects generically called
cluster in literature, see Sect 3.3 for discussion, only keeping
those with a firm Chandra detection of the intracluster medium
spatially coincident with a galaxy overdensity. The latter is an unambigous
detection of deep potential wells. By our choice, our
cluster sample is incomplete. However, our sample would also be
incomplete if every structure called cluster were included, 
because what is available today at very high redshift is 
a collection of objects, not a complete sample. 

The adoption of this criterium leads
to five $z>1.4$ clusters, listed in Table 1, out of the many ``cluster" detections
in the literature. Our list of $z>1.4$ clusters does not include the $z\sim1.5$
cluster by Tozzi et al. (2012) because the available Spitzer data sample 
only part of the cluster and are contaminated by an angularly nearby 
rich $z=1$ cluster\footnote{We note that 
Tozzi et al. (2012) report having
Spitzer [3.6] images with exposure time 10 times
shorter than the data available in the 
Spitzer archive. Our discovery of the
contamination mentioned below uses 970s long exposure, reduced and analyzed 
like all the other Spitzer data of this work. The
contamination has been discovered during the LF analysis.}.
Our list does not include the $z=1.62$ structure claimed by Tanaka et al.
(2012) to have an ICM detection because our re-analysis of the deep Chandra 
data available  does not confirm their ICM detection after flagging 4 arcsec
aperture regions centered on point sources.

The redshift range of bona fide $z>1.4$ clusters goes  from $z=1.41$ to
$z=1.8$, as detailed in Table 1.  All clusters but one have a redshift known
with two--digit (at least) precision. Instead, JKCS\,041 has a $\pm0.1$
redshift uncertainty, which is the least source of uncertainty in this work
because $m^*$ has a negligible change with redshift (i.e. a negligible $\partial m^* /
\partial z$) at high redshift. In passing, we note that the presence of a 
two-digit (spectroscopic) redshift does not necessarily indicate a more
precise redshift, as shown by the Gobat et al. (2011) group, whose initial
spectroscopic redshift has been lowered by $\delta z=0.08$ (Gobat 2012\footnote{http://www.sciops.esa.int/SYS/CONF2011/images/ \hfill \break
cluster2012Presentations/rgobat\_2012\_esac.pdf}), or
by MS\,1241.5+1710, whose initial spectroscopic  redshift has been increased
by $\delta z=0.237$ (Henry 2000).

All the five clusters have Spitzer observations. Table 1
lists the object ID (Column 1), the cluster redshift (Column 2),
and the Spitzer exposure time per pixel (Column 3). 

\begin{table}
\caption{The cluster sample.}
\begin{tabular}{l  l r }
\hline
ID  & $z$ & $t_{exp}$ \\
 &    & [s]   \\
\hline
JKCS\,041 	        & 1.8  & 1200 \\ 
IDCS\,J1426.5+3508      & 1.75 & 420 \\   
ISCS\,J1432.4+3250      & 1.49 & 1000 \\  
XMMXCS\,J2215.9-1738    & 1.45 & 1500 \\  
ISCS\,J1438.1+3414      & 1.41 & 1000 \\  
\hline  
\end{tabular}
\end{table}

\subsection{The data and analysis}

The basic data used in our analysis are 
the standard pipeline pBCD (post Basic Calibrated Data)
products delivered by the Spitzer Science Center (SSC).
These data include flat-field corrections, dark
subtraction, linearity, and flux calibrations. Additional steps
included pointing refinement, distortion correction, and mosaicking.
Cosmic rays were rejected during mosaicking by sigma-clipping. The
pBCD products do not merge observations taken in  different
astronomical observations requests (AORs). AORs are therefore mosaicked
together using SWARP (Bertin, unpublished), making use of the weight maps.
Images of IDCS\,J1426.5+3508 were already reduced
and distributed by Ashby et al. (2009) so we used them. For the two other 
clusters
at right ascension about $14^h$ we use the deeper data available
in the archive instead. 

Sources were detected using SExtractor (Bertin \& Arnout 1996), making
use of weight maps. Star/galaxy separation was performed by using the
stellarity index provided by SExtractor.
We conservatively kept a high posterior threshold ($class_{star}=0.95$),
rejecting ``sure star" only ($class_{star}>0.95$)
in order not to reject galaxies (by unduly putting them in the star class),
leaving some residual stellar contamination in
the sample. This contamination is later dealt with statistically, 
with background and foreground galaxies on the cluster line-of-sight.
The use of a high posterior threshold is very different from 
using a concentration index to select galaxies because the latter
at low signal-to-noise or for low-resolution
observations is susceptible to (mis)classifing galaxies as stars.
Objects brighter than 12.5 mag are often
saturated and therefore removed from the sample. No cluster galaxy is
that bright.

As pointed out by Ashby et al. (2009), images are already moderately crowed
with a 420 sec exposure, 
and this makes the catalog completeness much brighter than the limiting
depth (see also Mauduit et al. 2012). Because crowding is greather
in cluster regions, incompletness is more severe in these 
lines-of-sight. For example,
a circle of 1 arcmin radius centered on
IDCS\,J1426.5+3508 is 50 \% overdense compared to the average
line-of-sight. The overdensity is higher for richer clusters.
Completeness at [3.6]=19.5 mag is 50 \% in the general
field around IDCS\,J1426.5+3508 (in agreement with Ashby et al. 2009), 
but a few percentage points in the cluster (overdense) direction. 
If neglected, the differential (crowding) correction
biases the LF parameters (makes the cluster LF flatter than actually is,
and, via the parameter covariance, biases the characteristic magnitude, too). 
Furthermore, very low values of completeness do not allow reliable LFs
to be derived, even when crowding is accounted for by
following, for example, Andreon (2001). 
To avoid the danger
associated with an important (cluster-, radial- and magnitude-dependent) crowding
correction, we adopted quite bright magnitude limits, corresponding to 
90 \% completeness in average lines-of-sights and we ignored
fainter galaxies, as, e.g., in De Propris et al. (2007). The 90\% 
completeness is 
estimated by comparing galaxy counts in non-cluster lines-of-sight to 
crowding-corrected galaxy counts
taken from Barmby et al. (2008). We checked with the Ashby et al. (2009)
data that this procedure returns a completeness vs magnitude 
estimate that is very close to the one
derived by Ashby et al. (2009) from the recovery rate of simulated 
point sources
added in the images. Our threshold magnitudes are very conservative; 
for example, Mancone et al. (2010, 2012) used 50 \% completeness
magnitude in control field directions, which are about 2 mag fainter
than those we would have chosen.

\begin{figure*}
\centerline{%
\psfig{figure=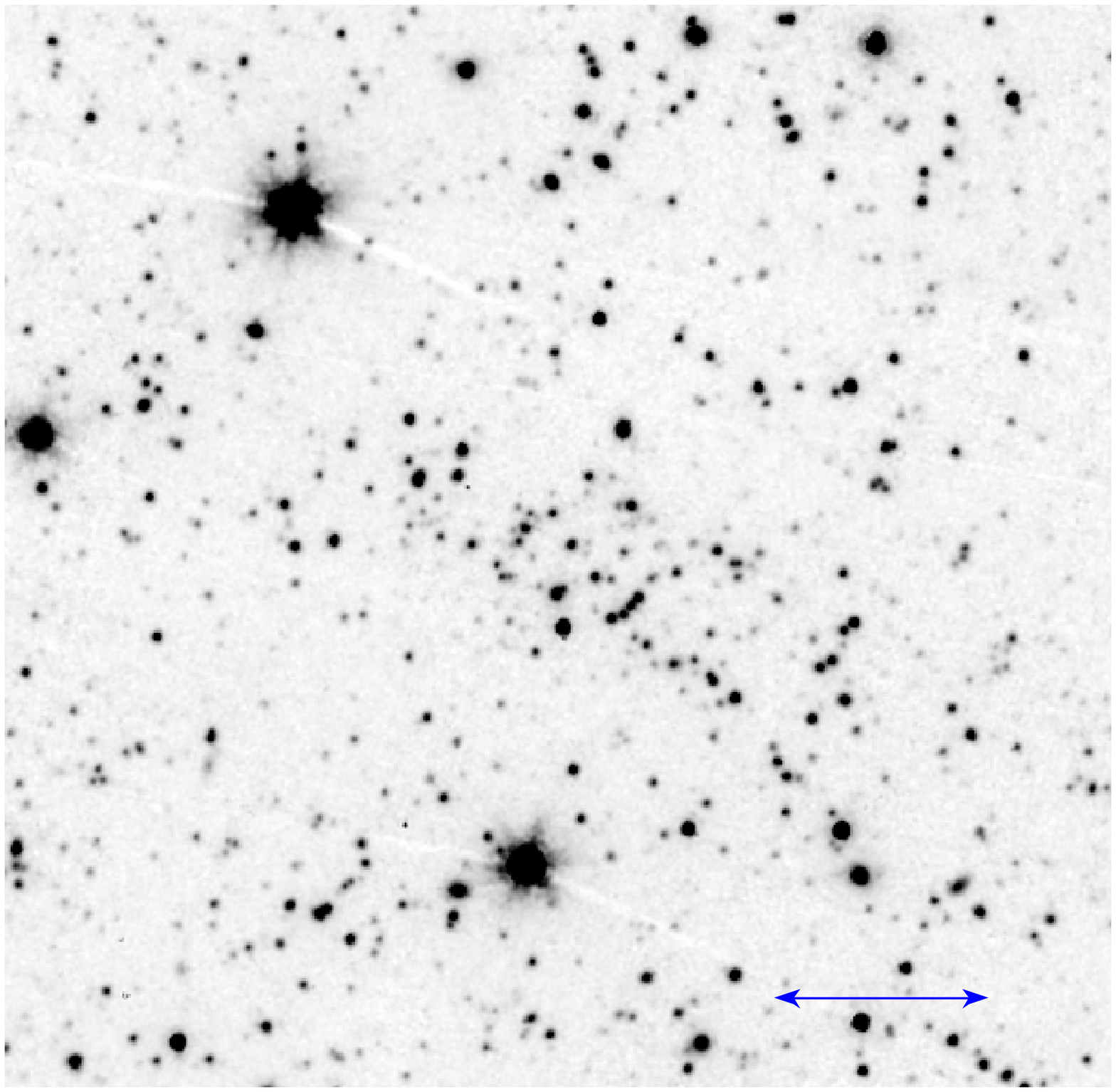,height=7truecm,clip=}
\psfig{figure=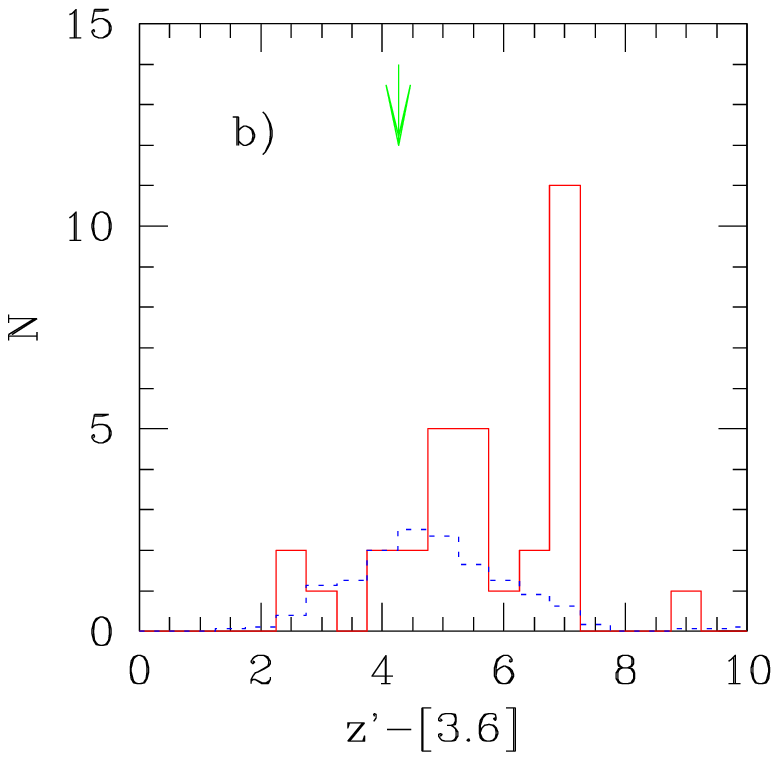,width=5truecm,clip=}
\psfig{figure=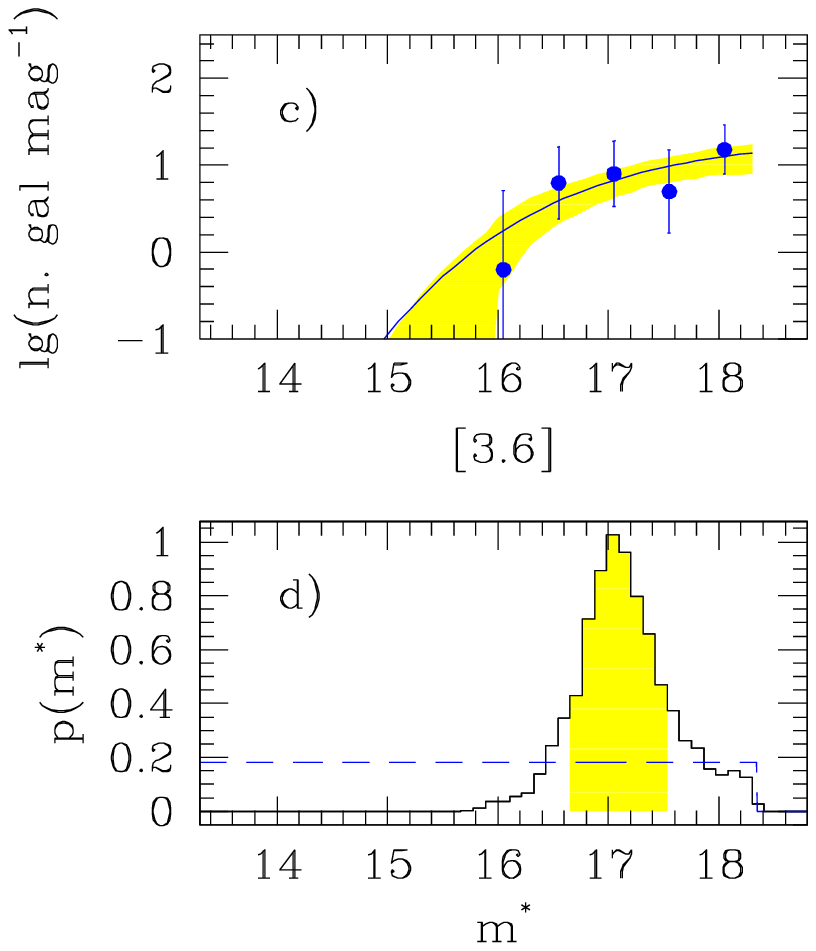,width=6truecm,clip=}}
\caption[h]{{\it Left Panel:} [3.6] image of JKCS\,041. The field of view is $5\times5$
arcmin, the ruler is 1 arcmin wide. North is up, east to the left. {\it Central Panel:} 
Color distribution in the cluster line-of-sight (solid red histogram) and in the
control field direction (blue dashed histogram). The arrow indicates the expected
color of a Grasil young star-forming galaxy (green arrow) 
at the cluster redshift. {\it Top--right panel:} Luminosity function. 
The points marks the LF as usually derived in
the astronomical literature, whereas the continuous (blue) curve
and the shaded area mark the LF and its 68 \% errors based on our Bayesian analysis. 
The fit is performed on unbinned data, not on the binned data plotted
in this figure. {\it Bottom--right panel:}
Probability distribution of $m^*$. The shortest 68 \% interval is
shaded. The dashed line indicates the assumed prior.
}
\label{fig:JKCS041}
\end{figure*}

To compute the luminosity function (LF), we adopted a Bayesian approach,
as done for other clusters (e.g. Andreon 2002, 2006, 2008, 2010, Andreon et al. 
2006,  2008a, Meyers et al. 2012). We accounted for the background
(galaxies in the cluster line-of-sight), estimated  
in adjacent lines-of-sight. We adopted
a Schechter (1976) luminosity function for cluster galaxies
and a third--order power law for the background distribution. 
The likelihood expression were taken from Andreon, Punzi \& Grado (2005), 
which is an
extension of the Sandage, Tammann \& Yahil (1979) likelihood expression  
for the case where a background is present. We adopted uniform priors for
all parameters, except for the faint end slope $\alpha$, taken
to be equal to $-1$ for comparison with lower redshift LF determinations.
In particular, we took a uniform prior on $m^*$ between 12.5 mag and our
adopted (quite bright, indeed) limiting mag. The Bayesian approach allowed 
us to easily 
propagate all uncertainties and their covariance into the 
luminosity function parameters and derived quantities.
As a visual check, we also
computed the luminosity function by binning galaxies in magnitude bins 
(e.g. Zwicky 1957, Oemler 1974, and many papers since then). These LFs are
plotted in figures 1 to 5 as points with errors. Our fit instead used 
unbinned galaxy counts.
For all clusters we adopted an aperture with a 1 arcmin radius (about 500
kpc).

To limit the contamination by galaxies in the cluster foreground, we
removed (with one exception discussed in next section) from the sample
all galaxies that are too blue to be at the cluster redshift, as in Andreon et al.
(2004) and other works (e.g. Mancone et al. 2012). We anticipated that
this choice has no impact on the results. For rich clusters, 
we also computed the LF of galaxies of all colors to check our assumption. 
To estimate the bluest acceptable
color a cluster galaxy  may plausibly have, we computed the  
$optical-[3.6]$ color (in the various adopted photometric indexes) 
of an exponentially increasing  star formation history (SFH) model, 
adopting the SFH of the template
named Sc in Grasil. In this way most of the stars at the cluster
redshift are newborn. We adopted a formation redshift $z=3$ (i.e. the
template has very young stellar populations, less than 0.8 to 2.3 Gyr old)
and a Salpeter initial mass function with lower/upper limit fixed to
0.15/120 $M_\odot$. Grasil (Silva et al. 1998; Panuzzo et al. 2005)
is a code to compute the spectral evolution of stellar systems 
the effects of dust taking into
account, which absorbs and scatters optical and UV
photons and emits in the IR-submm region.

Optical magnitudes  are derived from two sources: 
for JKCS\,041 and XMMXCS\,J2215.9-1738 we used
CFHTLS Deep z' bands (from
K-band detected WIRDS catalogs, Bielby et al. 2012).
For clusters at right ascension $\sim14^h$, 
we used NOAO Deep I band catalogs (Jannuzi \& Dey 1999).
The data are of adequate depth
for our purposes: we have one, or at most two, galaxies detected
at [3.6] and undetected in the optical band (because our adoption
of a bright magnitude cut in the [3.6] band). 
For these optically undetected galaxies, their undetection makes
their color redder than the blue
template and therefore these sources are kept in the sample.

Finally, by integrating the LF we derived the number of cluster 
galaxies brighter than
two limiting magnitudes: a) 18.0 mag, the brighter among
all chosen threshold magnitude values. 
This allows us to properly compare cluster richnesses if
$m^*$ does not evolve in the studied redshift range;
b) our bright limiting magnitude. This
gives the number of cluster galaxies actually fitted. 
We emphasize that our richnesses computed above accounts for the existence of
background galaxies and for errors on, and fluctuation of, the background 
counts, 
as well as uncertainties derived from having sampled a finite, usually small, 
number of cluster galaxies.

\begin{figure*}
\centerline{%
\psfig{figure=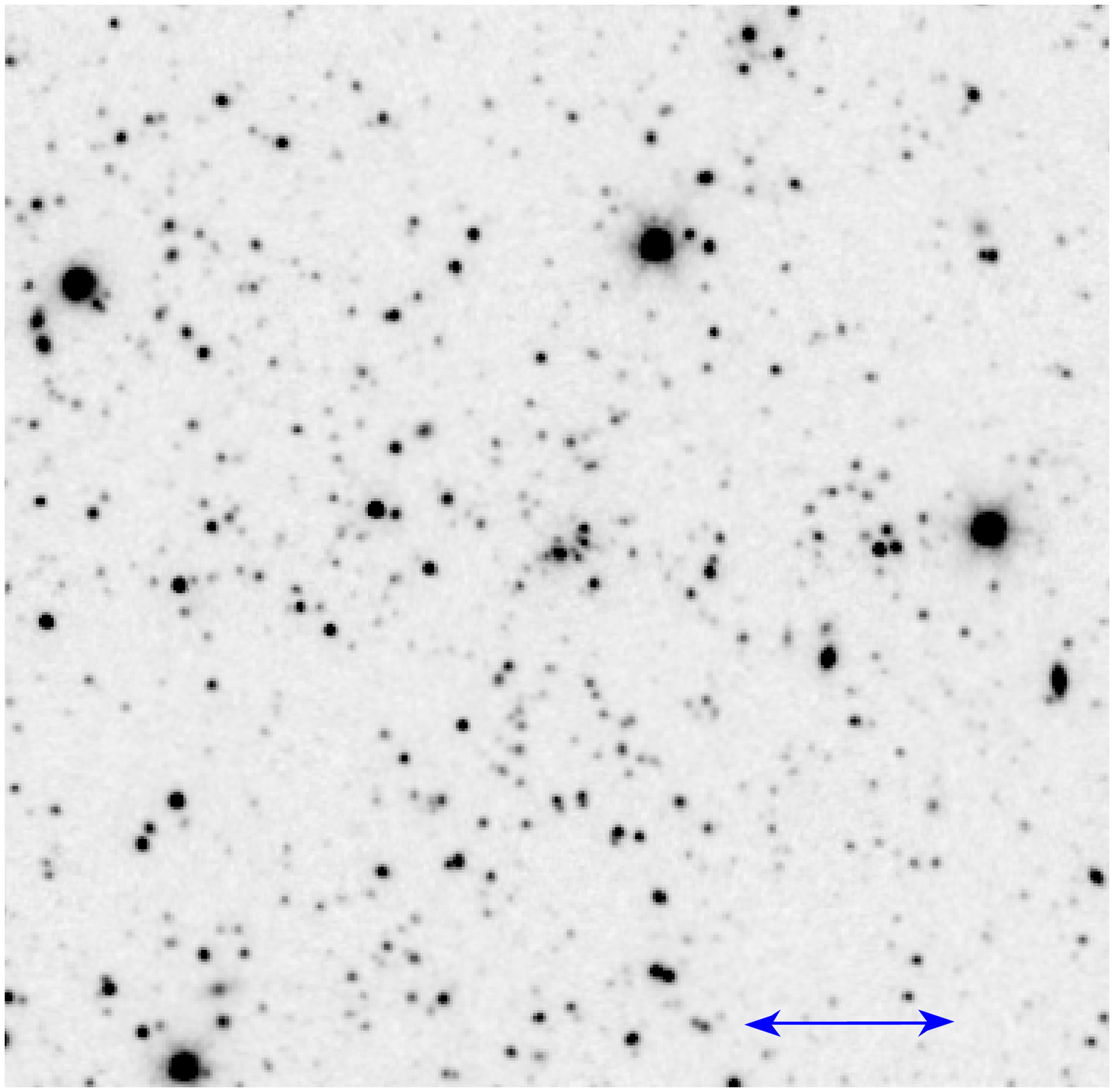,height=7truecm,clip=}
\psfig{figure=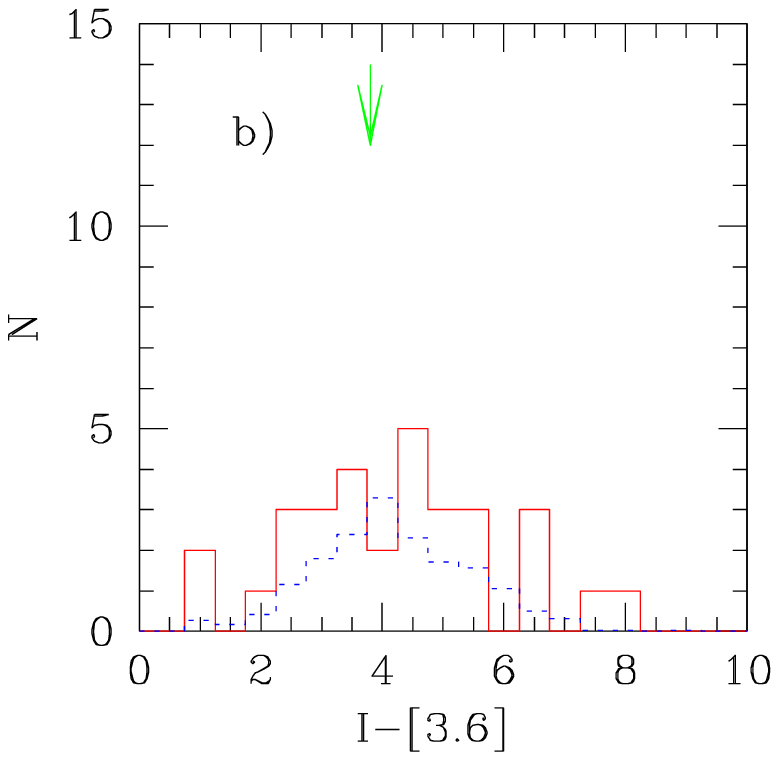,height=5truecm,clip=}
\psfig{figure=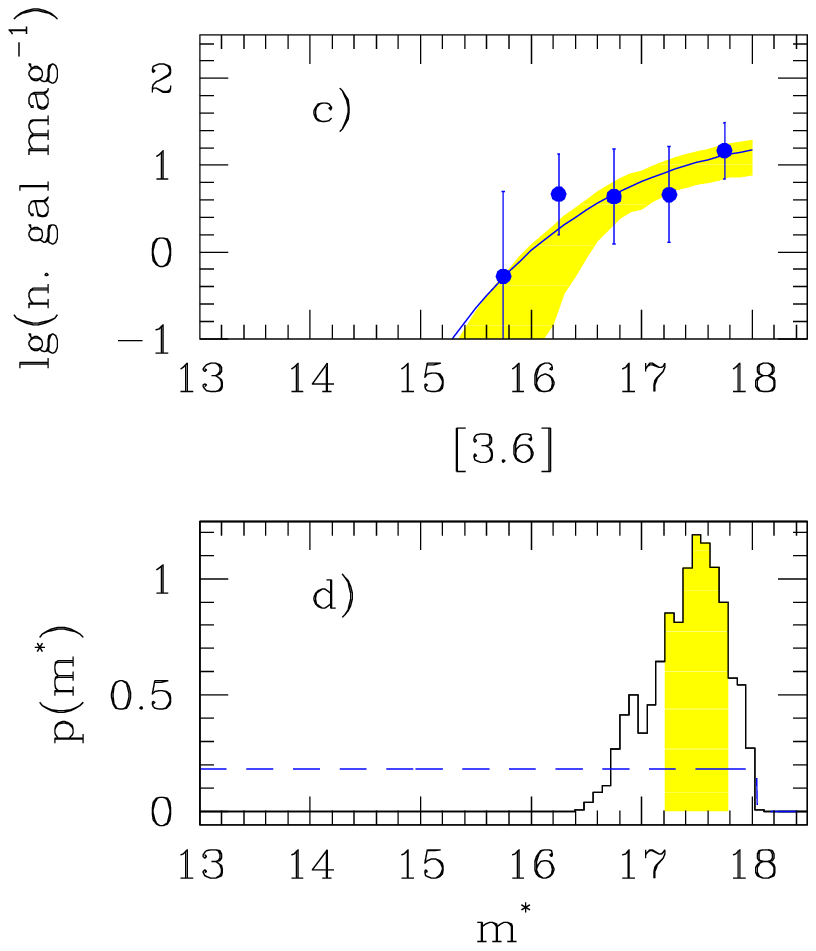,width=6truecm,clip=}
}
\caption[h]{Same as Fig.~\ref{fig:JKCS041}, but for the cluster IDCS\,J1426.5+3508.
}
\label{fig:IDCS}
\end{figure*}

\begin{figure*}
\centerline{%
\psfig{figure=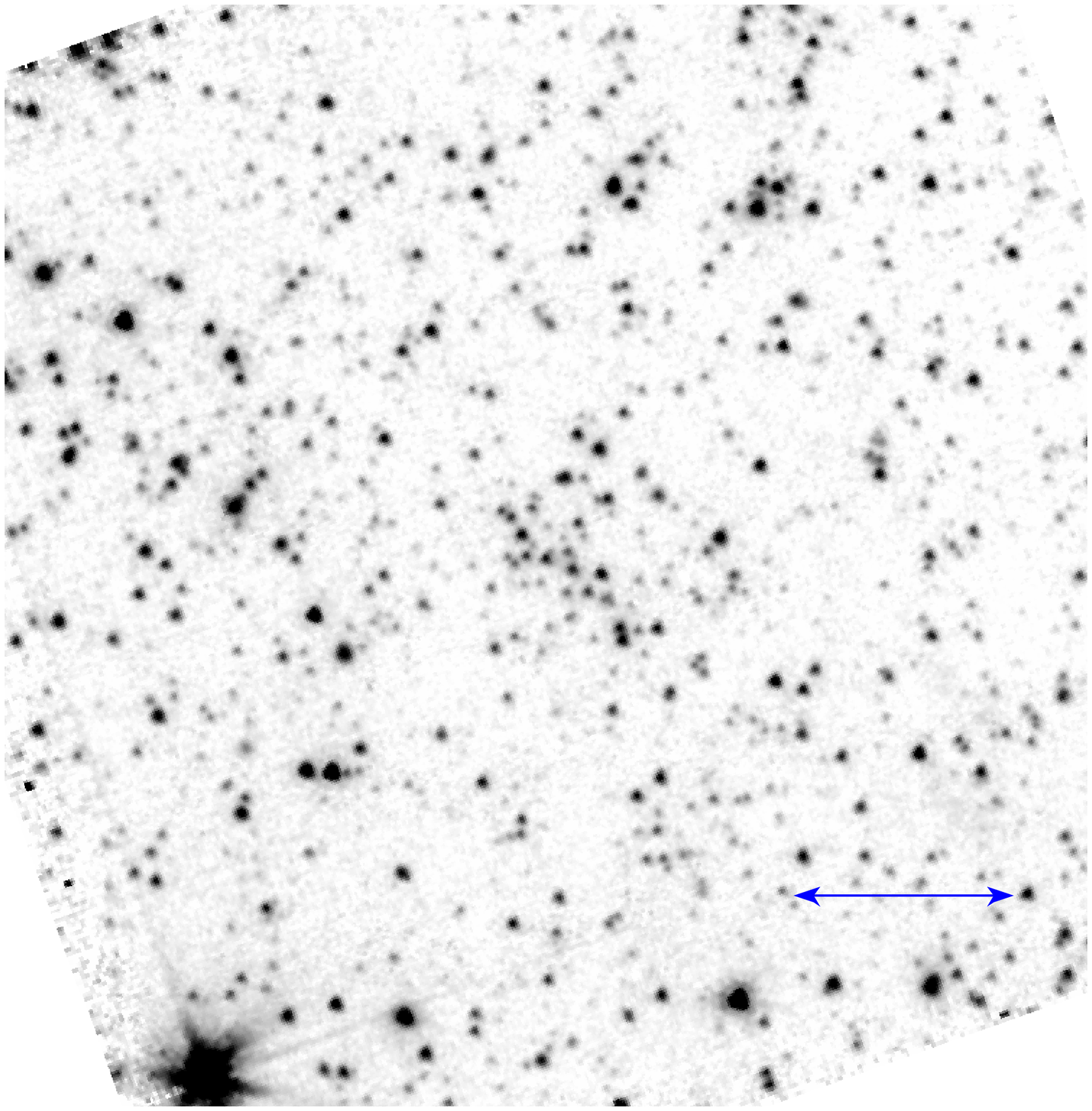,height=7truecm,clip=}
\psfig{figure=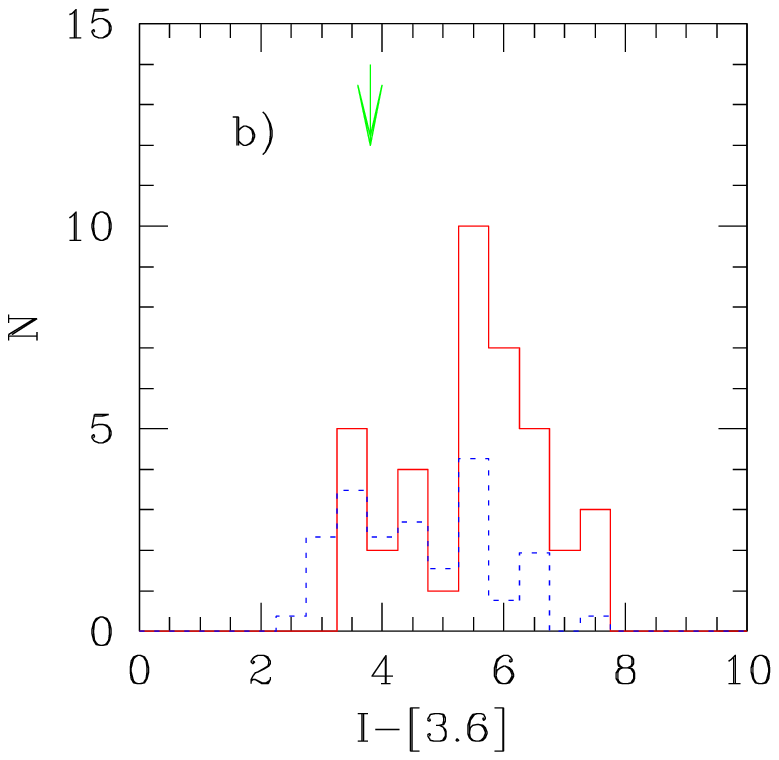,height=5truecm,clip=}
\psfig{figure=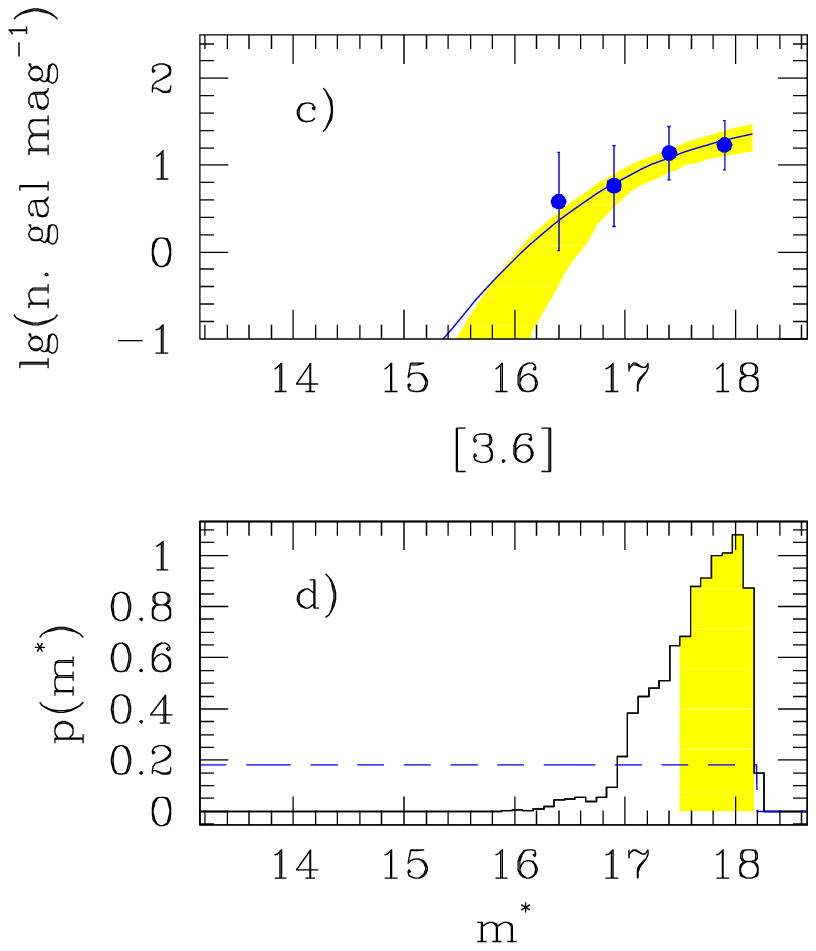,width=6truecm,clip=}
}
\caption[h]{Same as Fig.~\ref{fig:JKCS041}, but for the cluster ISCS\,J1432.4+3250.
}
\label{fig:z149}
\end{figure*}

\section{Results}

\subsection{Results for individual clusters}

\subsubsection{JKCS\,041 ($z\sim1.8$)}

JKCS\,041 (Andreon et al. 2009) stands out quite clearly as remarkable galaxy overdensity
in the left--hand panel of Figure~\ref{fig:JKCS041}. It
is the most distant cluster of our sample and also one of the most 
studied ones thanks to the deep data at various wavelengths.
It has been studied in the context
of the SZ scaling relations (Culverhouse et al. 2010), and it has been used to
measure the evolution of the $L_X-T$ scaling relation (Andreon, Trinchieri, 
\& Pizzolato 2011). JKCS\,041 color-magnitude relation
has been measured in Andreon \& Huertas-Company (2011), whereas the
age spread of galaxies on the red sequence is studied in Andreon (2011b). 
The relation between star formation and environment  
is determined in Raichoor \& Andreon (2012a) and
the evolution of the quenching rate with redshift 
(previously known as Butcher-Oemler effect) in
Raichoor \& Andreon (2012b).

Because of the presence of a group in the cluster southeastern 
outskirts (Andreon
\& Huertas-Company 2011), in our analysis we exclude all the 
southeastern quadrant of
JKCS\,041\footnote{A similar contamination is also
present for other high redshift
clusters, which were not retained in our final sample studied here, 
such as the $z=1.393$ 1WGA\,J2235.3-2557 (Mullis et al. 2005)
and the $z\sim1.5$ CXOJ1415.2+3610 (Tozzi et al. 2012) clusters.}. The red--sequence
population also  shows up clearly in the z'-[3.6] band (see the central panel of 
Fig.~\ref{fig:JKCS041} and compare with the other clusters). The color
distribution seems to indicate the presence of a blue population
at z'-[3.6] $\approx 5$ mag, so
redder than the blue spectrophotometric template (vertical
arrow), already pointed out using other filters in Raichoor \& Andreon (2012a).
There is no evidence of an excess of
galaxies bluer than the blue template (i.e. left of the vertical arrow
in the central panel of Figure~\ref{fig:JKCS041}).

The luminosity function of those $18\pm 5$ galaxies brighter than $18.3$ mag
in the three quadrants is shown in the top right--hand panel of Fig.~\ref{fig:JKCS041}. 
The characteristic magnitude is $17.0\pm0.5$ mag (see the bottom right--hand panel
for the probability distribution of it). The cluster has a richness
of $19\pm6$ galaxies, after accounting for the unused southeastern quadrant.

The luminosity function of red--sequence (i.e. $6.5<z'-[3.6]<7.5$ mag)
galaxies has an identical characteristic magnitude, $16.9\pm0.6$ mag,
as expected because most JKCS\,041 galaxies are on the red sequence.

\subsubsection{IDCS\,J1426.5+3508 ($z=1.75$)} 

As also shown in the left--hand panel of Figure~\ref{fig:IDCS}, 
IDCS\,J1426.5+3508 has a dense core of 
galaxies, whose clean detection obliged us to remove the image filtering when
detecting its galaxies using Sextractor.
The cluster has been detected, as JKCS\,041 was, as a galaxy overdensity 
(Stanford et al. 2011). Shallow Chandra data (Stanford et al. 2011)
allow the ICM to be detected, 
but not to be characterized (e.g. to estimate its temperature). The ICM is detected in 
absorption
via the Sunyaev-Zeldovich effect by Brodwin et al. (2012). The cluster displays
a red sequence (Stanford et al. 2011), not visible 
in the central panel of Figure~\ref{fig:IDCS}, but which shows up, 
at $I-[3.6]\sim6$ mag when 
considering fainter galaxies (our catalog is incomplete at these faint
magnitudes, however). Unique among the
five clusters considered in this paper, IDCS\,J1426.5+3508 shows
a possible presence of galaxies that are bluer than the blue 
spectrophotometric template (central panel). 
For this reason, the LF computation of IDCS\,J1426.5+3508 uses galaxies 
of all colors.

\begin{figure*}
\centerline{%
\psfig{figure=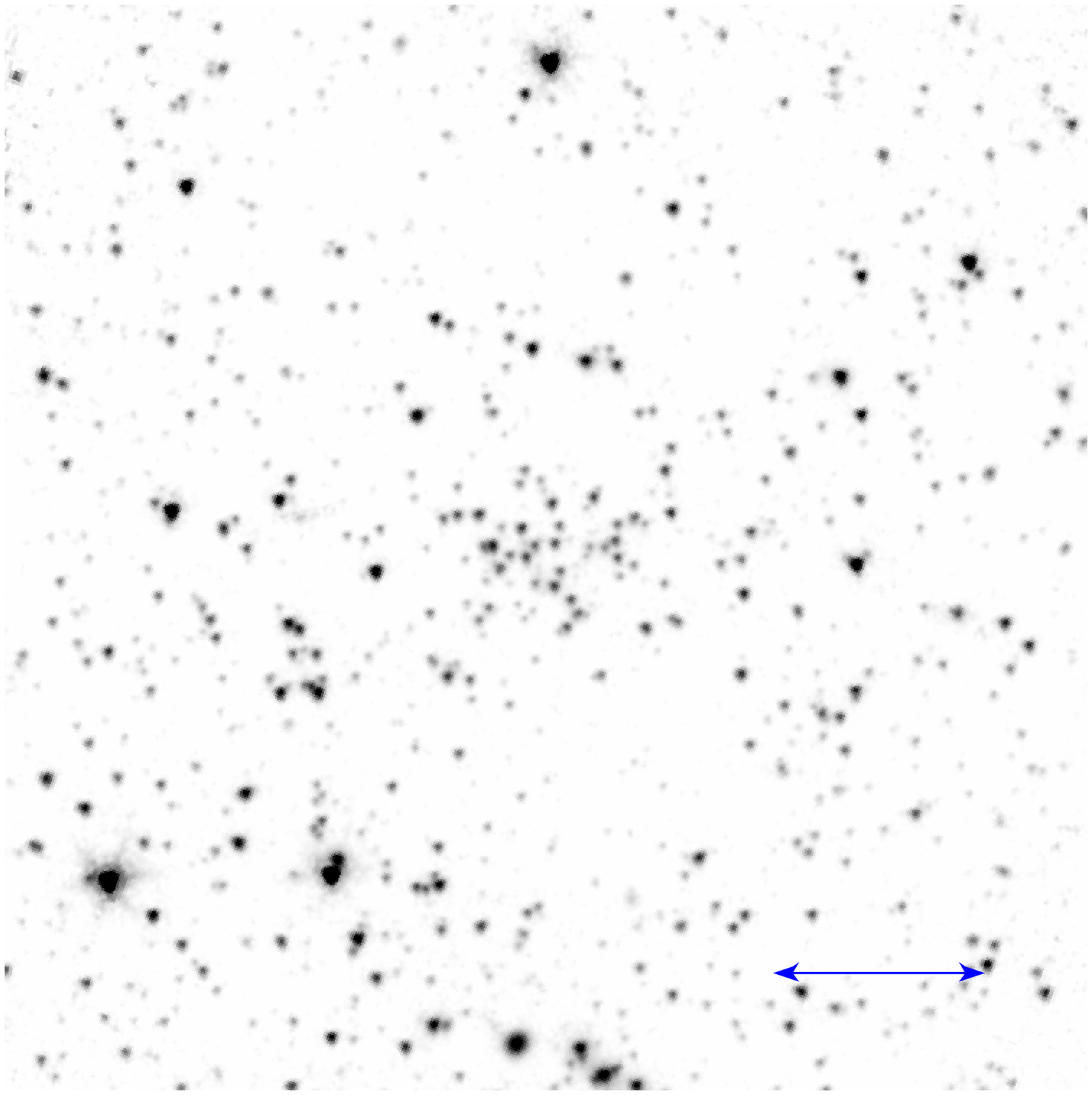,height=7truecm,clip=}
\psfig{figure=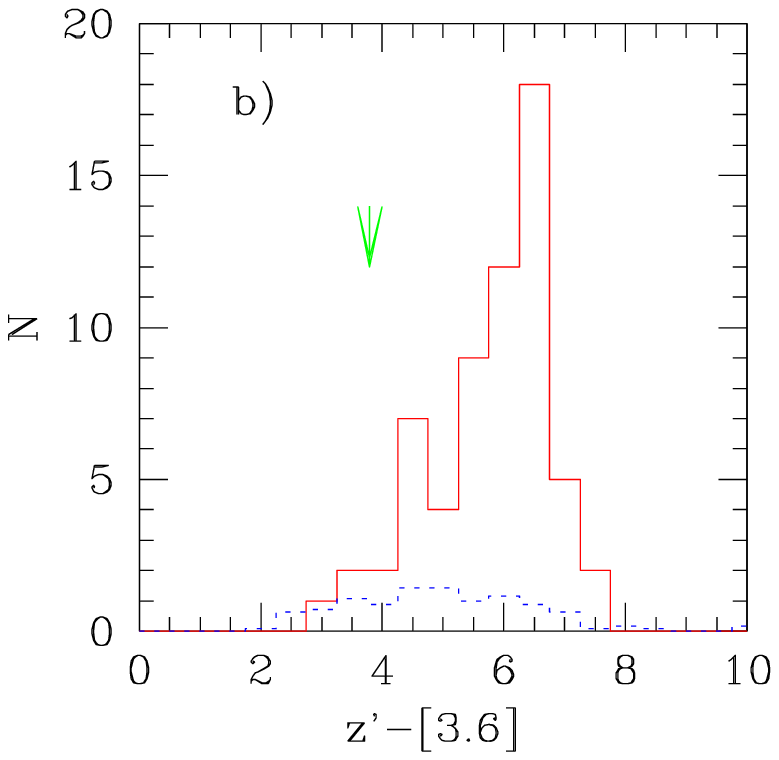,height=5truecm,clip=}
\psfig{figure=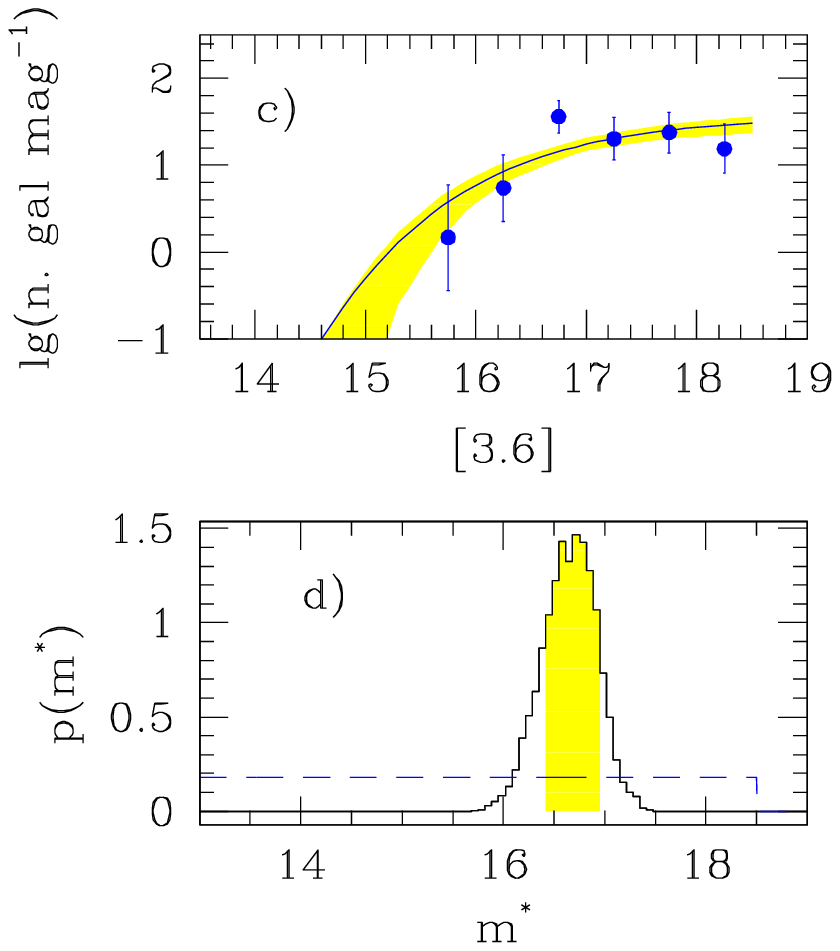,width=6truecm,clip=}
}
\caption[h]{Same as Fig.~\ref{fig:JKCS041}, but for the cluster XMMXCS\,J2215.9-1738.
}
\label{fig:cl2215}
\end{figure*}

\begin{figure*}
\centerline{%
\psfig{figure=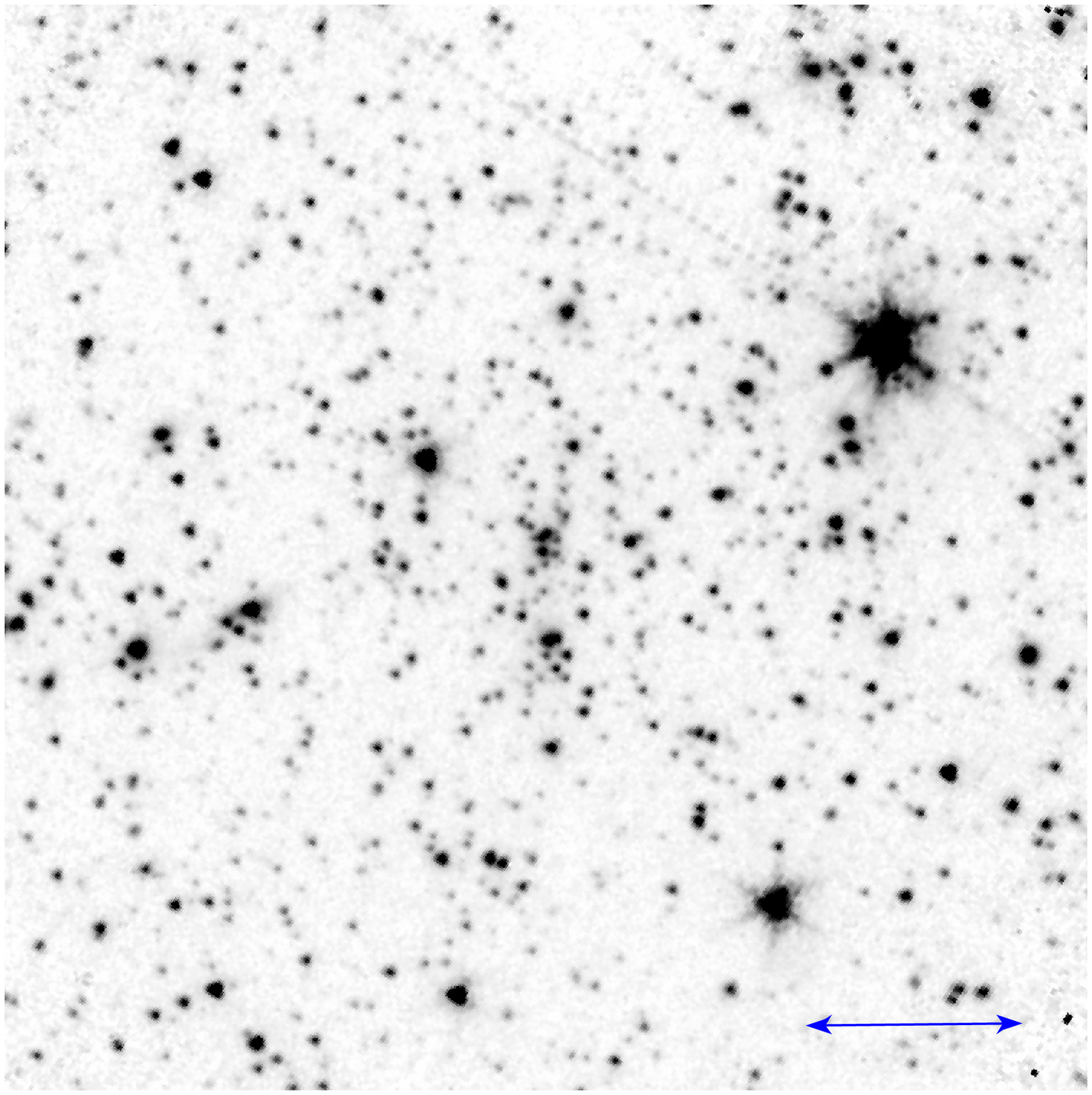,height=7truecm,clip=}
\psfig{figure=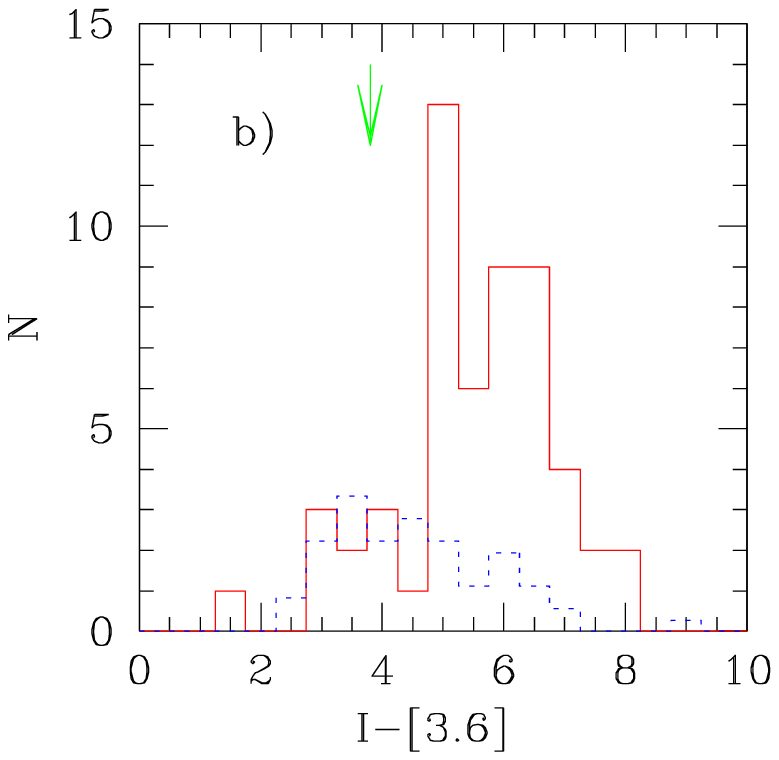,height=5truecm,clip=}
\psfig{figure=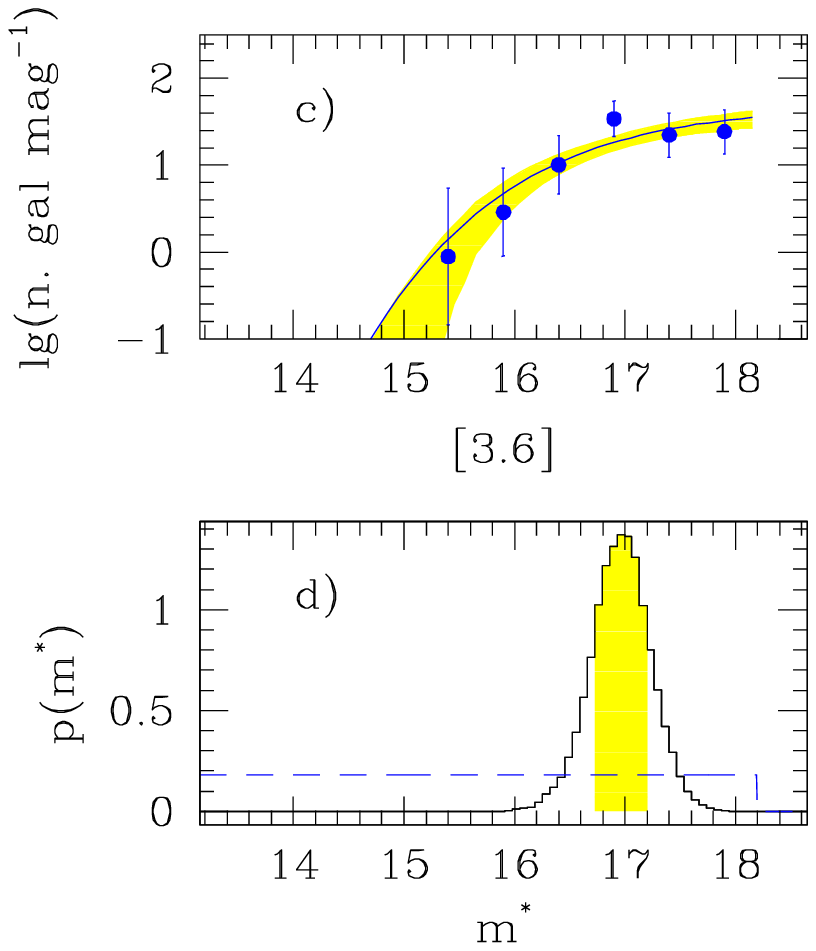,width=6truecm,clip=}
}
\caption[h]{Same as Fig.~\ref{fig:JKCS041}, but for the cluster ISCS\,J1438.1+3414.
}
\label{fig:z141}
\end{figure*}

The luminosity function of the $15\pm 6$ galaxies brighter than $18.0$ mag
is shown in the top right--hand panel. We find a characteristic magnitude of
$17.3\pm0.4$ mag, but we notice that the error amplitude depends on the adopted
prior (bottom-right panel): data allow lower values of $m^*$, although
with low probability, but the prior ($m^*<18.0$ mag) 
discard them. This situation occurs because the data used are too
shallow to bound the lower end of the $m^*$ probability
distribution of this cluster.
Indeed, IDCS\,J1426.5+3508 data are the shallowest in
our sample (see exposure times in Table 1).

IDCS\,J1426.5+3508 and JKCS\,041 have comparable richnesses ($15\pm 6$ vs
$19\pm6$ galaxies), although with different color distributions (JKCS\,041
has a larger fraction of red galaxies).

\subsubsection{ISCS\,J1432.4+3250 ($z=1.49$)} 

ISCS\,J1432.4+3250 (left--hand panel of Figure~\ref{fig:z149}) has, similar to
the previous two clusters,
been detected as a galaxy overdensity (Brodwin et al. 2011). In terms of spatial
distribution, it does not have a compact core of galaxies as 
IDCS\,J1426.5+3508 has.  Shallow Chandra
data, presented in Brodwin et al. (2011), allow the detection of the ICM, but
not its characterization. 
The cluster red sequence is studied
in Snyder et al. (2012) and also shows up at $I-[3.6]\sim6$ mag 
(see the central panel of Figure~\ref{fig:z149}). Similar to JKCS\,041 and
unlike IDCS\,J1426.5+3508, there is no evidence of a galaxy population bluer than
the blue spectrophotometric template (see the central panel).

The luminosity function of the $21\pm 6$ galaxies brighter than $18.15$ mag
is shown in the top right--hand panel. We find a characteristic magnitude of 
$17.4\pm0.4$ mag,
but we notice that the error amplitude depends on the adopted prior (bottom-right
panel), as for IDCS\,J1426.5+3508.

ISCS\,J1432.4+3250 has comparable richnesses ($17\pm 5$) to the two
clusters at higher redshift.

\subsubsection{XMMXCS\,J2215.9-1738 ($z=1.45$)} 

XMMXCS\,J2215.9-1738 (left panel of Figure~\ref{fig:cl2215}) is the most
distant X-ray selected cluster (Stanford et al. 2006). The cluster, initially
discovered with XMM-Newton,
has been re-observed with Chandra (Hilton et al. 2010), and the original XMM
detection was found to be heavily contaminated by point sources (already
suspected in XMM discovery data by Stanford et al. 2006). Although X-ray 
selected, hence likely overbright for its mass and temperature 
(Andreon,  Trinchieri,  \& Pizzolato 2011), the cluster turns
out to be underluminous
for its temperature assuming a self-similar evolution
(Hilton et al. 2010, Andreon,  Trinchieri, \& Pizzolato 2011).
Its color-magnitude relation is studied in Hilton et al. (2009)
and Meyers et al. (2012).

The cluster's red sequence shows up at $z'-[3.6]\sim6.5$ mag 
(see the central panel of figure \ref{fig:cl2215}). Unlike 
IDCS\,J1426.5+3508, there is no evidence of a galaxy population bluer than
the blue spectrophotometric template (see the central panel).
The luminosity function of the $49\pm 8$ galaxies brighter than $18.5$ mag
is shown in the top right--hand panel. We find a characteristic magnitude of 
$16.6\pm0.3$ mag (see the bottom right--hand panel
for the probability distribution of it).
ISCS\,J1432.4+3250 is richer than the other clusters at higher redshift ($36\pm 6$ 
vs $15$ to $20$ galaxies).

Richnesses and characteristic magnitude of the luminosity function 
derived removing the color selection 
are indistinguishable from those just derived (see Table~2)
because of the large dominance of red galaxies.

This cluster, at the spectroscopic redshift $z=1.45$, has a red
sequence 0.5 mag bluer than JKCS\,041, independently confirming that
JKCS\,041 has $z>1.45$ and that therefore JKCS\,041 has to be kept in 
the sample of
$z>1.4$ clusters\footnote{After the paper acceptance, JKCS\,041 has been
spectroscopic confirmed to be at high redshift by mean of HST spectroscopy.}. 
We emphasize that this color comparison
uses homogeneous photometry that is uniformly reduced 
($z'$ band by Bielby et al. 2010, 
this work for [3.6] band).

\subsubsection{ISCS\,J1438.1+3414 ($z=1.41$)} 

ISCS\,J1438.1+3414 (left--hand panel of Figure~\ref{fig:z141}) has  been detected as a
galaxy overdensity (Stanford et al. 2005). Deep Chandra data (Andreon,  Trinchieri,
\& Pizzolato 2011) allowed characterizing the ICM and measuring the evolution of the
$L_X-T$ scaling relation (Andreon,  Trinchieri,  \& Pizzolato 2011). Its red
sequence has been studied in Meyers et al. (2012). The cluster red sequence stands
out (see the central panel of Figure~\ref{fig:z141}), although it seems quite
broad. However, to accurately measure the width of the red sequence, 
a more tailored measurement of color is needed. 
Unlike IDCS\,J1426.5+3508, there is no evidence of a galaxy
population bluer than the blue template (see the central panel). 

The luminosity function of the $47\pm 8$ galaxies brighter than $18.15$ mag
is shown in the top right--hand panel. We find a characteristic magnitude of $16.8\pm0.3$ mag.
The cluster is quite rich, $42\pm7$ galaxies brighter
than 18 mag, as rich as XMMXCS\,J2215.9-1738, and much richer
than the clusters at higher redshift. 
Richnesses and characteristic magnitude of the luminosity function 
derived by removing the color selection 
are indistinguishable from those derived above (see Table~2)
because of the strong dominance of red galaxies.

\begin{table*}
\caption{Results}
\begin{tabular}{l l l l l l}
\hline
ID & $m^*$ & $n(<18.0)$ & ref mag & $n(<$ ref mag$)$ & Notes \\
(1) & (2) & (3) & (4) & (5)  \\
\hline
JKCS\,041 	        & $17.0\pm0.5$ & $19\pm6$ &  18.30 & $18\pm5$ & $z'-[3.6]> 4.3$ mag\\ 
IDCS\,J1426.5+3508      & $17.3\pm0.4$ & $15\pm6$ &  18.00 & $15\pm6$ & galaxies of all colors \\
ISCS\,J1432.4+3250      & $17.4\pm0.4$ & $17\pm5$ &  18.15 & $21\pm6$ & $I-[3.6]> 3.8$ mag \\
XMMXCS\,J2215.9-1738    & $16.6\pm0.3$ & $37\pm6$ &  18.50 & $52\pm8$ & $z'-[3.6]> 3.8$ mag\\
ISCS\,J1438.1+3414      & $16.8\pm0.3$ & $42\pm7$ &  18.15 & $47\pm8$ & $I-[3.6]> 3.8$ mag\\  
\hline
Other fits & & & \\
JKCS\,041 	        & $16.9\pm0.6$ & $13\pm4$ & 18.30 & $12\pm4$  & $6.5<z'-[3.6]< 7.5$ mag\\ 
XMMXCS\,J2215.9-1738    & $16.5\pm0.3$ & $38\pm6$ & 18.50 & $52\pm8$ & galaxies of all colors\\
ISCS\,J1438.1+3414      & $16.8\pm0.3$ & $43\pm8$ & 18.15 & $47\pm9$ & galaxies of all colors\\  
\hline
\end{tabular}
\hfill \break  
Richnesses in col (5) of JKCS\,041 refers to measurement in three
quarters of the cluster area, see text.
\end{table*}

\subsection{Collective analysis}

\begin{figure}
\centerline{\psfig{figure=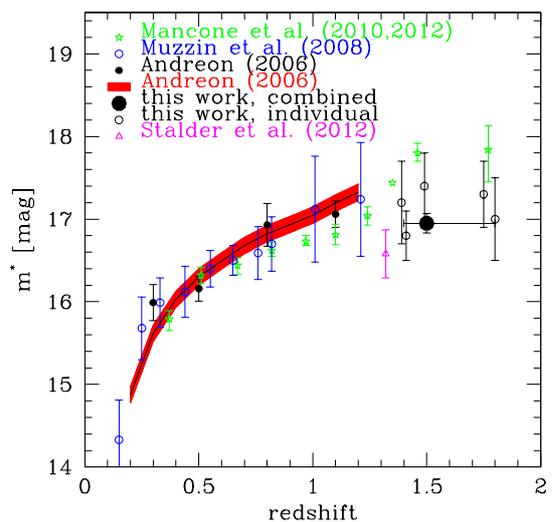,height=7truecm,clip=}}
\caption[h]{Characteristic magnitude vs redshift: observed values from
Andreon (2006, two derivations shown as solid small points and as shading),
Muzzin et al. (2008, open blue points, $z<1.3$), Stalder et al. (2012), and
Mancone et al. (2010, 2012). Error bar are not easy to compare across
works because they do not include the same terms in the error budget.} 
\label{fig:mstarz}
\end{figure}

\begin{figure}
\centerline{\psfig{figure=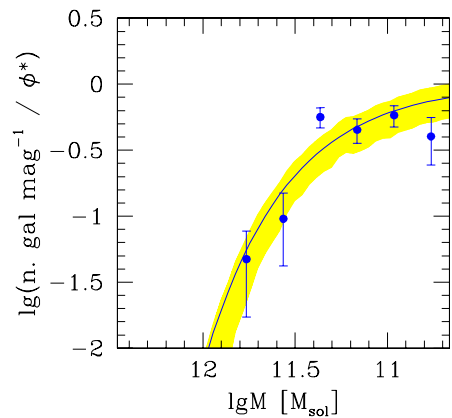,height=6truecm,clip=}}
\caption[h]{Combined mass function of the six $z>1.4$ clusters. Data points 
and error bars are computed as usual (e.g. Oemler 1973, see sec. 2.2).
The solid line is the mean model, the shaded region indicates the
68\% uncertainty of the model. 
}
\label{fig:lfall}
\end{figure}

Figure~\ref{fig:mstarz} shows the derived $m^*$ values of
the five $z>1.4$ clusters, as well as
previous determinations from the literature in the [3.6]
band (Andreon 2006; Muzzin et al.
2008; Mancone et al. 2010, 2012; Stalder et al. 2012), 
corrected in the case of Muzzin et al. (2008) for
the different faint end slope adopted. Andreon (2006), Muzzin et al.
(2008), and Mancone et al. (2010)
fitted, as we also do, an LF with a fixed slope.
The plotted errors are 
heterogeneous and not easily comparable
because they do not include the same terms in the error budget.
In particular, the Mancone et al. (2010) results are based on a few spectroscopically 
confirmed clusters and many putative cluster detections, including false ones,
as mentioned by the authors.  
At $z<1.2$, Muzzin et al. (2008) is consistent with the more precise
determination of Andreon (2006), whereas Mancone et al. (2010)
$m^*$ values display a slower change with redshift than the
other data. The difference is statistically significant if one assumes that 
the Mancone et al. (2010) errors are correctly estimated.

The five $z>1.4$ clusters studied in this paper (black open dots) have
$m^*$ values that are consistent among themselves. We therefore
combined the data of the five clusters by multiplying their likelihoods after
tying the characteristic magnitude parameters.  
We find $m^* = 16.92\pm0.13$ at $z=1.5$, the median
redshift of the six clusters (solid dot), 
based on the 150 member galaxies of the combined sample. This value is 
is 0.8 mag brighter than the values measured by Mancone et al. (2010)
in the same redshift range. The difference is statistically significance
if one assumes that the Mancone et al. (2010) errors are correctly estimated. 
As detailed in 
the technical appendix, 
we identify the likely reason for the disagreement, Mancone et al. (2010)
did not adopted the likelihood expression appropriate for the data used,  
and we also
exclude the possibility that the difference is due to an intrinsic variance of 
$M^*$.

We emphasize that in this combined
fit, the precise redshift of the five clusters is irrelevant, provided
they have $z>1.4$. Therefore, the lack of a spectroscopic redshift
for JKCS\,041 (or any other cluster) is not detrimental for the 
collective $m^*$ determination
(or for individual $m^*$, as it is self--evident). 
We also emphasize
that the highest redshift cluster, JKCS\,041, only contributes 10 \% of
the total number of galaxies, and that  an indistinguishable
result would therefore be found by dropping it from the sample.

Figure~\ref{fig:lfall} plots the galaxy mass function of the combined
five clusters. Mass is defined as the
integral of the star formation rate, and it is
derived from the [3.6] luminosity
assuming a single stellar population (SSP) formed at $z_f=2.5$, modeled by 
the 2007 version of Bruzual \& Charlot (2003) spectrophotometric population
synthesis code with solar metallicity
and a Chabrier (2003) initial mass function.  
Based on the 150 member galaxies of the combined sample,
the mass function is
determined with a 40 \% error in the $10.5<lgM<12$ M$_\odot$ range. 
The characteristic mass is $lgM^*=11.30\pm 0.05$ 
M$_\odot$, where the error does not account for the systematic errors
coming from the conversion from luminosity to mass (e.g. about the stellar
initial mass function shape). 

\begin{figure}
\centerline{\psfig{figure=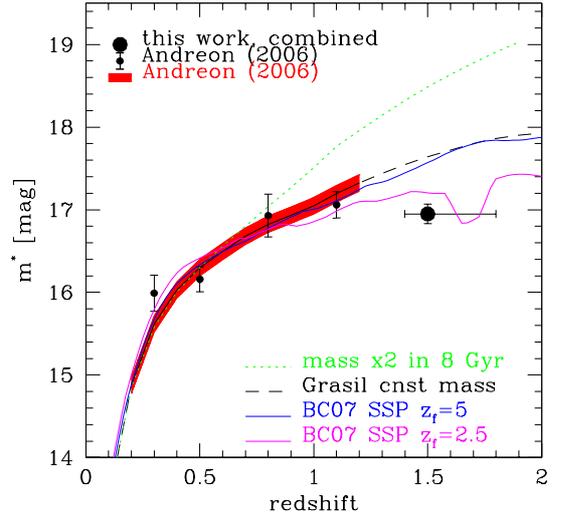,height=7truecm,clip=}}
\caption[h]{Characteristic magnitude vs redshift, with several model
predictions.} 
\label{fig:mstarz_mod}
\end{figure}

Figure~\ref{fig:mstarz_mod} summarizes the observational data by keeping
$m^*$ determinations limited to confirmed clusters 
(Muzzin et al. 2008 and Mancone et al. 2010 studied a mix of real clusters
and cluster detections) and
setting the Mancone et al. (2012) and Stadler et al. (2012)
determinations aside because these works use galaxy counts in a magnitude regime 
where a crowding correction is, at best, compelling, but ignored (see also
the technical appendix). The characteristic
luminosity $m^*$ has the same value at $1.4<z<1.8$ as at $z\sim1$. This is
the main result of this work, which still holds true when keeping all $z<1.3$ data
(see Fig.~\ref{fig:mstarz}).

Figure~\ref{fig:mstarz_mod} also plots
the luminosity evolution expected for several mass--growth histories. The very
steep luminosity evolution drawn in the figure (dotted curve) is a Grasil 
model in which
the mass doubled in the last 8 Gyr, first explained and then
ruled out in Andreon (2006).
Mass-evolving models are rejected by the data, because a mass
growth makes galaxies brighter (because more massive) at lower redshift, and 
thus these models do not fit the data. For example, exponentially
declining $\tau$ models are rejected by the data for
all $\tau$ tau values higher than 0.5 Gyr, 
the precise value depending however on the
adopted $z_f$. Models involving
mergers do not fit the data 
because the latter makes descending galaxies more massive and therefore
brighter, i.e. moves $m^*$ to brighter
values going to lower redshift, while the points at ``low" ($z\sim1$)
redshift are too faint for the $1.4<z<1.8$ point(s). 

Therefore, from now on we only  consider models with no ongoing (in the last
11 Gyr) mass growth. The solid line
shows the luminosity evolution of an SSP, modeled as previous SSP but formed at 
$z_f=5$. The dashed line close to this SSP is a Grasil non-evolving 
mass model (named E both in the Grasil package and in
Andreon 2006). This model is a bit more physical than Bruzual \& Charlot (2003)
because, for example, a metallicity evolution is allowed during the 1 Gyr long
star formation episode started at $z=5$ (all other Grasil parameters are kept
to the default value). These models are rejected too by the data, being
too faint at very high redshift (or too bright at lower redshift
if models are normalized at $1.4<z<1.8$).
To have an overall flat evolution at $z>>1$ and the observed trend
at $z\la 1$, one needs to boost the light
at $1.4<z<1.8$ without increasing the cluster mass, otherwise the descending
galaxies would be too massive (bright) for the low ($z\la 1$) redshift
data points. This can be achieved by considering an SSP 
with a formation redshift very close to the highest observed redshift, 
$z_f=2.5$, a mere 0.5 Gyr age difference from the age of the
most distant cluster in the sample  or 0.9 Gyr from the age
of the second most distant cluster. To boost the light 
observed at $1.4<z<1.8$, and thus emitted
at early ages (1 to 2 Gyr), it is necessary to
use the 2007 version of the Bruzual \& Charlot (2003)
models or Maraston (2005) and to keep a low $z_f$. In fact, the 2003 release of
Bruzual \& Charlot (2003) gives galaxies that are too faint (for the data) 
galaxies in the critical 1 to 2 Gyr age because of
the different treatment of the thermally--pulsing asymptotic 
giant branch (TP-AGB). Larger $z_f$ does not boost the light enough
to reproduce the data:
the $z_f=5$ case is dipicted in the figure and does not fit the data. 
Similarly, a $z_f=3$ SSP does not fit the data.

To summarize, models with mergers or recent star formation 
(in the past
11 Gyr) are rejected because they are too
luminous at mid-to-high $z$ (or, equivalently, too faint at very 
hight redshift). Populations that are too
old (SSPs with $z_f=5$) are rejected 
because too faint at very high $z$.
The similarity of $m^*$ at $1.4<z<1.8$ and $z\sim1$
implies an assembly time earlier than $z=1.8$ and, at the same
time, a star formation episode that is not much earlier than $z_f=2.5$ in order
to boost the luminosity of the galaxies observed between 0.9 to 2.0 Gyr
after it (our $1.4<z<1.8$ galaxies). 

Past works dealing with luminosity/mass functions
usually give a {\it lower} limit on the formation redshift or on the
last star--formation episode. The 
large redshift baseline and the robustly measured $m^*$ values of this work
are able to distinguish SSPs with $z_f=2.5$ 
or $z_f=5$ (in Fig.~\ref{fig:mstarz_mod} 
these SSPs are hardly distinguishable below 
$z\sim1.2$ and widely different at higher redshift), and are also able to 
indicate that the {\it upper} limit of the latest star formation episode 
did not preceed the redshift of the two most distant clusters in the sample
by a long time, only about 0.9 Gyr.

We emphasize that exponentially declining models with small $\tau$ (e.g. 0.1
Gyr) show a luminosity evolution very similar to SSP models, and thus are an
equally acceptable fit to the data. Similarly, models with twice the 
solar metallicity
are hardly distinguishable in Fig.~\ref{fig:mstarz_mod} from solar
metallicity models, so are also acceptable. However, adopting
them does not change our conclusion. Finally, we note
that the impact of TP-AGB stars on galaxy spectral energy distributions
is still under discussion (Zibetti et al. 2011): the lower their contribution
to the overall emission in the rest-frame near--infrared band is,
the harder it is to fit our data.

\subsection{Cluster choice}

To be included in this work, a $z>1.4$ structure should have
a firm,  Chandra, ICM detection that is
spatially coincident with a galaxy overdensity. This may seem  overly
restrictive, and in fact
at an early stage
of this work we considered other possible criteria, but we found
them unsuitable. 

Initially, we accepted $z>1.4$ spectroscopic confirmed
clusters. However, relying on the common
``spectroscopic confirmation" to name a cluster is dangerous
at high redshift: the presence of a 
spectroscopically confirmed cluster-sized galaxy overdensity 
is not a guarantee of the presence of a cluster. It is enough
to think about the zoo
of structures with different names (proto-clusters, redshift spikes), often
with several concordant redshifts in areas of a few Mpc$^2$. The usual spectroscopic criteria
have been shown to have low reliability by Gal et al. (2008)
using a real spectroscopic survey. This is exemplified by the
spectroscopic confirmation  
of the Gobat et al. (2011) $z=2.07$ group: recent spectroscopic data 
(Gobat et al.
2012\footnote{http://www.sciops.esa.int/SYS/CONF2011/images/ \hfill \break
cluster2012Presentations/rgobat\_2012\_esac.pdf}) 
show that none of their original 11
spectroscopic members used to spectroscopically confirm the group
belong to the group, all being more than $8000$ km s$^{-1}$
away from the group. 
Given the danger of the ``spectroscopic confirmation", 
we choose to require a firm ICM detection. 

Later, we considered a suitable definition of cluster to be every
galaxy overdensity with a spatially coincindent X-ray detection,
either from Chandra or XMM. However,
a weak XMM detection, typical of most clusters in the XMM Deep Cluster
Survey (Fassbender 2011) and of other searches, such as Henry et al.
(2010), is an 
ambigous cluster detection: it may be ICM emission 
or an AGN misclassified as extended source 
because of the low XMM resolution and the low
signal--to--noise.
This is the case of the Papovich et al. (2010) structure, initially named
cluster, and renamed proto-cluster in Papovich et al. (2012). It was detected
as a $4\sigma$ XMM source (as many Fassbender 2011 ``clusters"). Later  
pointed Chandra follow-up observations revealed a bright point source and
no extended emission (Pierre et al. 2012) using
observations planned to measure the intracluster medium temperature to better than
30 \% out to $r_{500}$ (M. Pierre Chandra proposal abstract). 
Similarly, 
the Gobat et al. (2011) group with a $3.5\sigma$ XMM detection after point source
subtraction is undetected in deep Chandra observations (Gobat et al. 2011).

Therefore, to avoid the risk of computing the luminosity function of an
environment mis-identified as a cluster, i.e. of comparing ``apples" (at high redshift)
to ``oranges" at low redshift (Andreon \& Ettori 1999), we asked for a firm, 
Chandra, 
ICM detection that is spatially coincident with a galaxy overdensity.
The superb Chandra angular resolution allows 
X-ray point
sources (1" wide) to be easily recognized  
from extended (20" wide) ICM emission, even in the low
signal--to--noise regime, unlike XMM.

\section{Summary}

We analyzed deep Spitzer data of the five $z>1.4$ clusters
with a firm detection of the ICM spatially coincident with
a galaxy overdensity. This definition of cluster
gets rid of the many sorts of cluster-sized structures known at high
redshift (such as proto-cluster), which may differ from
clusters and allows us to be certain we are comparing ``apples to apples"
(Andreon \& Ettori 1999).
The analyzed data are deep (about 1000s), but to avoid
a cluster- radial- magnitude-dependent, unreliable crowding correction,
we only consider bright galaxies (brighter than 18.0-18.5 mag), about
2 mag brighter than other authors would choose for data with the same exposure
time. 
The four clusters differ in richness (ISCS\,J1438.1+3414 and 
XMMXCS\,J2215.9-1738 are twice richer than ISCS\,J1432.4+3250, 
IDCS\,J1426.5+3508 and JKCS\,041) and morphological appareance.
By adopting the correct expression of the likelihood for the data in hand
we derived the luminosity function in the [3.6] band  and 
the characteristic magnitude $m^*$, the latter by marginalizing over the 
remaining
parameters except $\alpha$. 
Since the five $m^*$ values are found to be consistent 
with each other, we combined the data in
the unique statistically acceptable way, by multiplying the likelihood of each
individual determination. We found a characteristic luminosity of
$m^* = 16.92\pm0.13$ at $z=1.5$, the median
redshift of the six clusters. Assuming a luminosity-to-mass conversion
fixed by an SSP with $z_f=2.5$,
we found a characteristic mass $lgM^*=11.30\pm 0.05$ at $z=1.5$ and a mass function
determined with about 40 \% error in the $10.5<lgM<12$ M$_\odot$ range
from the 150 member galaxies of the combined sample. 

We found
that the characteristic luminosity and mass does not evolve 
between $z\sim1$ and $1.4<z<1.8$, directly ruling out ongoing mass
assembly between these epochs because massive galaxies are already
present at $z=1.8$. Lower redshift build--up epochs were
already ruled out by previous works, leaving only $z>1.8$ as a possible
epoch for the mass build--up. 
Populations that are  too
old (SSPs with $z_f=5$) are rejected 
because they are too faint at very high $z$.
The observed values of $m^*$ at very
high redshift are, however, too bright for galaxies without any
star formation shortly preceeding the observed redshift.
The similarity of $m^*$ at $1.4<z<1.8$ and $z\sim1$
implies a star formation episode no earlier than $z_f=2.5$ in order
to boost the luminosity of the galaxies observed at $1.4<z<1.8$
without increasing their mass. For the first time, mass/luminosity
functions are able to robustly distinguish tiny differences between
formation redshifts ($z_f=2.5$ from $z_f=3$) and to set
{\it upper} limits to the last star formation episode. This 
did not preceed the redshift of the two most distant clusters in the sample
by a long time, only about 0.9 Gyr. In short, $1.4<z<1.8$
is the post starforming age of massive cluster galaxies, we 
found that massive cluster galaxies were still 
forming stars at $z\sim2.5$ and that they did not grow in mass 
at later times.

\begin{acknowledgements} 
This work is based on observations made with the Spitzer Space 
Telescope.

\end{acknowledgements}

\appendix

\section{Technical addedum about the LF determination}

Methods for deriving the luminosity function date back to Zwicky (1957) at
least, with newer methods usually more properly addressing  the complicate
features of the astronomical data not considered by previous methods (see
Johnston 2011 for a detailed list of references and Andreon 2011a for a listing
of many of the awkward features of the astronomical data). For challenging
estimations such as those of high redshift clusters, these ``complications"
include the use of the correct likelihood expression for the handled data and,
in the case of deep data, the (radial- magnitude- cluster- dependent) crowding
corrections  or the adoption of a bright limiting magnitude. If the objects
used to build the LF include putative clusters (i.e. spurious or false
cluster detections) or objects of a different nature from clusters (e.g.
filaments), this uncertainty should be folded into the likelihood expression.
The same is true if clusters have a photometric redshift, unless $m^*$ is
constant in the redshift uncertainty range. If galaxies are selected with
photometric redshift or their photometric redshift is used in the LF
determination, uncertainties (both statistics and systematics) should be folded
into the likelihood expression, too. Both effects introduce a bias on $m^*$ if
neglected, and this can be easily appreciated by remembering that contamination
and photometric redshift errors work as a convolution filter making the LF
broader, thus biasing $m^*$ even in the simplest case (symmetric
errors). Complicated bias patterns are introduced when asymmetry is important. 
If galaxies without optical counterparts are excluded from the LF computation,
the likelihood expression should be modified to correct for the bias induced by
the forced optical detection requirement. 

Mancone et al.
(2010, 2012) faced most of these issues but did not adopt the
likelihood appropriate for the data used.
This is the reason, in our opinion, for the (formally
statistically) different $m^*$ change with redshift at $z<1.2$ between Mancone et al.
(2010) and the other works, as well as for the  (formally statistically) different
$m^*$ values between Mancone et al. (2010) and this work at
$z\sim 1.5$.
Before interpreting them as genuine differences, due for example to  
two populations of clusters with widely different $m^*$ values, with
Mancone et al. (2010) primarly sampling one population and this work
the other, one should first
make certain that all determinations are robustly derived and
refer to clusters as we usually intend them (the author consider
sheets, filaments, proto-clusters, and false cluster detections 
as fairly different from clusters), and second,  
three of the five clusters studied in this work 
are likely also in the Mancone 
et al. (2010) sample (using shallower data, however).

For completeness, we also explored whether the difference 
between our $m^*$ values and those in Mancone et al. (2010) at $z\sim1.5$
may be due to an intrinsic variance of $M^*$ values.  By performing a Monte
Carlo simulation, we computed the probability that
if $M^*$ has an intrinsic  scatter, five out five of the 
clusters studied in this work are all within 0.6 mag, and the
$m^*$ of the combined cluster is 0.85 mag brighter (or more)
than the Mancone et al. (2010) value.
To this aim, as prior probability distribution for the intrinsic 
(i.e. accounting for measurement errors)
scatter at $z\sim1.5$ we adopted the
posterior probability distribution computed for the  
17 clusters at $0.29<z<1.06$ in Andreon et al. (2004), whose luminosity
functions have been fitted by holding $\alpha$ fixed, as we and
Mancone et al. (2010) both do. 
This distribution may be approximated by a normal distribution centered
on $0.02$ mag, with sigma $0.22$ mag and truncated at zero. 
The distribution is quite broad,
meaning that we allow the possibility in our simulations 
that the instrinsic scatter may be very large. In fact, 5\% of our
simulations have an intrinsic $M^*$ scatter larger than $0.4$ mag. Note
that in simulations we adopted the actual probability distribution, 
not the Normal approximation mentioned above. Then, we generated 
60000 simulated data sets, each one composed of five clusters, 
with $m^*$ values having the same errors as our observed values and
having Mancone et al. (2010) mean $m^*$ (and
intrinsic scatter as detailed above). Finally, we counted
how many times we observe mean $m^*$ offsets in the simulated data larger than
those in the real data (0.85 mag) and individual $m^*$ values
all within 0.6 mag (the observed maximal
difference between the individual $m^*$ measured by us).
We found no case in 60000 simulations, 
i.e. the probability  of observing a larger
disagreement because of the intrinsic variance in $M^*$ is a
negligible $2\ 10^{-5}$. 

To allow a possible evolution of
the intrinsic scatter between the redshift where it is measured,
$0.3<z<1.1$, and the redshift where we need to known its value,
$z\sim1.5$, we performed other simulations 
assuming that at higher redshift the intrinsic scatter is 
twice higher (if we reduce the scatter at higher redshift, it becomes
even more implausible to observe the observed $m^*$ offset). 
Also in this case,  we found no case matching the observations
out of our 60000 simulations.
In short, it is very unlikely that
the intrinsic scatter on $M^*$ is the source of the disagreement between
$z\sim1.5$ determinations in these two works.  

We emphasize that our search for a physical reason that explains
differences in the mean $m^*$  
assumes that measurements and errors (i.e. the likelihood) are (is) correct
whereas those in one of the compared works is not. Therefore,
our search, performed at the request of the referee, should not be interpreted 
as indication that we believe that the observed differences of the mean $m^*$ 
is genuine.


\begin{thebibliography}{}

\bibitem[Andreon(2001)]{2001ApJ...547..623A} 
Andreon, S.\ 2001, ApJ, 547, 623 

\bibitem[Andreon(2002)]{2002A&A...382..495A} 
Andreon, S.\ 2002, A\&A, 382, 495 

\bibitem[Andreon(2006)]{2006MNRAS.369..969A} 
Andreon, S.\ 2006, MNRAS, 369, 969 


\bibitem[Andreon(2010)]{2010MNRAS.407..263A} 
Andreon, S.\ 2010, MNRAS, 407, 263 (A10)

\bibitem[Andreon(2011)]{2011arXiv1112.3652A} 
Andreon, S.\ 2011a, in Astrostatistical Challenges for the New Astronomy,
ed. J. Hilbe, Springer Series on Astrostatistics (arXiv:1112.3652) 

\bibitem[Andreon(2011)]{2011A&A...529L...5A} 
Andreon, S.\ 2011b, A\&A, 529, L5 


\bibitem[Andreon \& Ettori(1999)]{1999ApJ...516..647A} 
Andreon, S., \& Ettori, S.\ 1999, ApJ, 516, 647 

\bibitem[Andreon \& Huertas-Company(2011)]{2011A&A...526A..11A} 
Andreon, S., \& Huertas-Company, M.\ 2011, A\&A, 526, A11 


\bibitem[]{} 
Andreon, S., Punzi, G., Grado, A., 2005, MNRAS, 360, 727

\bibitem[Andreon et al.(2004)]{2004MNRAS.353..353A} Andreon, S., Willis, 
J., Quintana, H., et al.\ 2004, \mnras, 353, 353 

\bibitem[Andreon et al.(2006)]{2006MNRAS.372...60A} 
Andreon, S., Cuillandre, J.-C., Puddu, E., \& Mellier, Y.\ 2006, 
        MNRAS, 372, 60


\bibitem[Andreon et al.(2008)]{2008MNRAS.385..979A} 
Andreon, S., Puddu, E., de Propris, R., \& Cuillandre, J.-C.\ 2008a, 
        MNRAS, 385, 979 


\bibitem[Andreon et  al.(2009)]{2009A&A...507..147A} 
Andreon, S., Maughan, B., Trinchieri, G., \& Kurk, J.\ 2009, A\&A, 507, 147 


\bibitem[Andreon et al.(2011)]{2011MNRAS.412.2391A} 
Andreon, S., Trinchieri, G., \& Pizzolato, F.\ 2011, MNRAS, 412, 2391 

\bibitem[Ashby et al.(2009)]{2009ApJ...701..428A} 
Ashby, M.~L.~N., Stern, D., Brodwin, M., et al.\ 2009, ApJ, 701, 428 

\bibitem[Barmby et al.(2008)]{2008ApJS..177..431B} 
Barmby, P., Huang, J.-S., Ashby, M.~L.~N., et al.\ 2008, ApJS, 177, 431 

\bibitem[Bertin \& Arnouts(1996)]{1996A&AS..117..393B} 
Bertin, E., \& Arnouts, S.\ 1996, A\&AS, 117, 393 

\bibitem[Bielby et al.(2012)]{2012A&A...545A..23B} 
Bielby, R., Hudelot, P., McCracken, H.~J., et al.\ 2012, A\&A, 545, A23 

\bibitem[Brodwin et al.(2011)]{2011ApJ...732...33B} 
Brodwin, M., Stern, D., Vikhlinin, A., et al.\ 2011, ApJ, 732, 33 

\bibitem[Brodwin et al.(2012)]{2012ApJ...753..162B} 
Brodwin, M., Gonzalez,  A.~H., Stanford, S.~A., et al.\ 2012, ApJ, 753, 162 

\bibitem[Bruzual \& Charlot(2003)]{2003MNRAS.344.1000B} 
Bruzual, G., \& Charlot, S.\ 2003, MNRAS, 344, 1000 

\bibitem[Chabrier(2003)]{2003PASP..115..763C} 
Chabrier, G.\ 2003, PASP, 115, 763 

\bibitem[Culverhouse et al.(2010)]{2010ApJ...723L..78C} 
Culverhouse, T.~L.,  Bonamente, M., Bulbul, E., et al.\ 2010, ApJ, 723, L78 


\bibitem[de Propris et al.(1999)]{1999AJ....118..719D} 
De Propris, R., Stanford, S.~A., Eisenhardt, P.~R., Dickinson, M., \& Elston, R.\ 1999, AJ, 118, 719 

\bibitem[De Propris et al.(2007)]{2007AJ....133.2209D} 
De Propris, R., Stanford, S.~A., Eisenhardt, P.~R., Holden, B.~P., 
	\& Rosati, P.\ 2007, AJ, 133, 2209 


\bibitem[Fassbender et al.(2011)]{2011NJPh...13l5014F} 
Fassbender, R., B{\"o}hringer, H., Nastasi, A., et al.\ 2011, 
	New Journal of Physics, 13, 125014 

\bibitem[Gal et al.(2008)]{2008ApJ...684..933G} 
Gal, R.~R., Lemaux, B.~C., Lubin, L.~M., Kocevski, D., \& Squires, G.~K.\ 2008, ApJ, 684, 933 

\bibitem[Gobat et al.(2011)]{2011A&A...526A.133G} 
Gobat, R., Daddi, E., Onodera, M., et al.\ 2011, A\&A, 526, A133 

\bibitem[Henry et al.(2010)]{2010ApJ...725..615H} 
Henry, J.~P., Salvato,  M., Finoguenov, A., et al.\ 2010, ApJ, 725, 615 

\bibitem[Mancone et al.(2010)]{2010ApJ...720..284M} 
Mancone, C.~L., Gonzalez, A.~H., Brodwin, M., et al.\ 2010, ApJ, 720, 284 

\bibitem[Mancone et al.(2012)]{2012ApJ...761..141M} 
Mancone, C.~L., Baker, T., Gonzalez, A.~H., et al.\ 2012, ApJ, 761, 141 

\bibitem[Maraston(2005)]{2005MNRAS.362..799M} 
Maraston, C.\ 2005, MNRAS, 362, 799 

\bibitem[Mauduit et al.(2012)]{2012PASP..124..714M} 
Mauduit, J.-C., Lacy, M., Farrah, D., et al.\ 2012, PASP, 124, 714 

\bibitem[Meyers et al.(2012)]{2012ApJ...750....1M} 
Meyers, J., Aldering, G., Barbary, K., et al.\ 2012, ApJ, 750, 1 


\bibitem[Mullis et al.(2005)]{2005ApJ...623L..85M} 
Mullis, C.~R., Rosati,  P., Lamer, G., et al.\ 2005, ApJ, 623, L85 

\bibitem[Muzzin et al.(2008)]{2008ApJ...686..966M} 
Muzzin, A., Wilson, G., Lacy, M., Yee, H.~K.~C., \& Stanford, S.~A.\ 2008, ApJ, 686, 966 

\bibitem[Jannuzi \& Dey(1999)]{1999ASPC..191..111J} 
Jannuzi, B.~T., \& Dey, A.\ 1999, Photometric Redshifts and the 
	Detection of High Redshift Galaxies, ASP Conference Series, Vol. 191, 
	Edited by R. Weymann, L. Storrie-Lombardi, M. Sawicki,
	and R. Brunner. ISBN: 158381-017-X, p. 111 



\bibitem[Johnston(2011)]{2011A&ARv..19...41J} 
Johnston, R.\ 2011, A\&A Review, 19, 41 

\bibitem[Henry(2000)]{2000ApJ...534..565H} 
Henry, J.~P.\ 2000, ApJ, 534, 565 

\bibitem[Hilton et al.(2010)]{2010ApJ...718..133H} 
Hilton, M., Lloyd-Davies, E., Stanford, S.~A., et al.\ 2010, ApJ, 718, 133 

\bibitem[Hilton et al.(2009)]{2009ApJ...697..436H} 
Hilton, M., Stanford,  S.~A., Stott, J.~P., et al.\ 2009, ApJ, 697, 436 

\bibitem[Oemler 1974]{1974ApJ...194....1O} 
Oemler, A., Jr. 1974, ApJ, 194, 1 

\bibitem[Panuzzo et al.(2005)]{2005astro.ph..1464P} 
Panuzzo, P., Silva, L., Granato, G.~L., Bressan, A., \& Vega, O.\ 2005, 
in "The Spectral Energy Distribution of Gas-Rich Galaxies: Confronting Models 
with Data", eds. C.C. Popescu and R.J. Tuffs, AIP Conf. Ser., 
in press, (astro-ph/0501464) 

\bibitem[Papovich et al.(2010)]{2010ApJ...716.1503P} 
Papovich, C., Momcheva, I., Willmer, C.~N.~A., et al.\ 2010, ApJ, 716, 1503 

\bibitem[Papovich et al.(2012)]{2012ApJ...750...93P} 
Papovich, C., Bassett, R., Lotz, J.~M., et al.\ 2012, ApJ, 750, 93 

\bibitem[Pierre et al.(2012)]{2012A&A...540A...4P} 
Pierre, M., Clerc, N., Maughan, B., et al.\ 2012, A\&A, 540, A4 

\bibitem[Raichoor \& Andreon(2012)]{2012A&A...543A..19R} 
Raichoor, A., \& Andreon, S.\ 2012a, A\&A, 543, A19 

\bibitem[Raichoor \& Andreon(2012)]{2012A&A...537A..88R} 
Raichoor, A., \& Andreon, S.\ 2012b, A\&A, 537, A88 

\bibitem[Sandage et al.(1979)]{1979ApJ...232..352S} 
Sandage, A., Tammann, G.~A., \& Yahil, A.\ 1979, ApJ, 232, 352 

\bibitem[Schechter(1976)]{1976ApJ...203..297S} 
Schechter, P.\ 1976, ApJ, 203, 297 

\bibitem[Silva et al.(1998)]{1998ApJ...509..103S} 
Silva, L., Granato,  G.~L., Bressan, A., \& Danese, L.\ 1998, ApJ, 509, 103 

\bibitem[Snyder et al.(2012)]{2012ApJ...756..114S} 
Snyder, G.~F., Brodwin,  M., Mancone, C.~M., et al.\ 2012, ApJ, 756, 114 

\bibitem[Stalder et al.(2012)]{2012arXiv1205.6478S} 
Stalder, B., Ruel, J., Suhada, R., et al.\ 2012, ApJ, submitted (arXiv:1205.6478) 

\bibitem[Stanford et al.(2005)]{2005ApJ...634L.129S} 
Stanford, S.~A.,  Eisenhardt, P.~R., Brodwin, M., et al.\ 2005, ApJ, 634, L129 


\bibitem[Stanford et al.(2006)]{2006ApJ...646L..13S} 
Stanford, S.~A., Romer, A.~K., Sabirli, K., et al.\ 2006, ApJL, 646, L13 

\bibitem[Stanford et al.(2012)]{2012ApJ...753..164S} 
Stanford, S.~A., Brodwin, M., Gonzalez, A.~H., et al.\ 2012, ApJ, 753, 164 

\bibitem[Strazzullo et al.(2010)]{2010A&A...524A..17S} 
Strazzullo, V., Rosati, P., Pannella, M., et al.\ 2010, A\&A, 524, A17 


\bibitem[Tanaka et al.(2012)]{2012arXiv1210.0302T} 
Tanaka, M., Finoguenov,  A., Mirkazemi, M., et al.\ 2012, PASJ, in press (arXiv:1210.0302) 

\bibitem[Tozzi et al.(2012)]{2012arXiv1212.2560T} 
Tozzi, P., Santos, J.~S., Nonino, M., et al.\ 2012, arXiv:1212.2560 


\bibitem[Zibetti et al.(2013)]{2013MNRAS.428.1479Z} 
Zibetti, S., Gallazzi, A., Charlot, S., Pierini, D., \& Pasquali, A.\ 2013, MNRAS, 428, 1479 

\bibitem[Zwicky(1957)]{1957moas.book.....Z} 
Zwicky, F.\ 1957, Morphological astronomy, Berlin: Springer


\end{thebibliography}
\end{document}